\documentclass{article}

\usepackage{arxiv}

\usepackage[utf8]{inputenc} 
\usepackage[T1]{fontenc}    
\usepackage{hyperref}       
\usepackage{url}            
\usepackage{booktabs}       
\usepackage{amsfonts}       
\usepackage{amsmath}
\usepackage{soul}
\usepackage{nicefrac}       
\usepackage{microtype}      
\usepackage{lipsum}
\usepackage{graphicx}
\graphicspath{ {./images/} }

\DeclareMathOperator*{\argmin}{arg\,min}

\title{Differentiable model-based adaptive optics for two-photon microscopy}

\author{
 Ivan Vishniakou$^*$ \\
  Center of Advanced European\\ Studies and Research (caesar)\\
  53175 Bonn, Germany
   \And
 Johannes D. Seelig\thanks{Corresponding authors: ivan.vishniakou@caesar.de, johannes.seelig@caesar.de} \\
  Center of Advanced European\\ Studies and Research (caesar)\\
  53175 Bonn, Germany\\
}

\begin{document}
\maketitle
\begin{abstract}
Aberrations limit scanning fluorescence microscopy when imaging in scattering materials such as biological tissue. Model-based approaches for adaptive optics take advantage of a computational model of the optical setup. Such models can be combined with the optimization techniques of machine learning frameworks to find aberration corrections, as was demonstrated for focusing a laser beam through aberrations onto a camera   \cite{vishniakou2020differentiable}.
Here, we extend this approach to two-photon scanning microscopy. 
The developed sensorless technique finds corrections for aberrations in scattering samples
and will be useful for a range of imaging application, for example in brain tissue.
\end{abstract}


\section{Introduction}
 
Using adaptive optics for imaging in aberrating samples requires finding appropriate corrections, which
can be determined using wavefront sensors \cite{kerr2008imaging, rodriguez2018adaptive, rotter2017light, yoon2020deep}. 
Alternatively, a variety of so called sensorless approaches, that don’t require a wavefront sensor, have been developed. 
One class of algorithms takes advantage of optimization. For example, iteratively modulating and updating an excitation wavefront depending on the resulting fluorescence intensity can be used for finding aberration corrections
\cite{vellekoop2015feedback, rotter2017light}.

Other approaches for adaptive optics additionally take advantage of prior information about the optical system by including system information in a computational model \cite{gonsalves2014perspectives, jefferies2002sensing, hanser2004phase, song2010model, linhai2011wavefront, yang2015model, antonello2015modal}. In this case, optimization can be used for finding aberrations which enter the model as undetermined parameters.
Different from the above mentioned optimization approaches, all data is already provided at the beginning of the optimization process and not recorded iteratively \cite{yang2015model}.

More recently, the development of machine learning frameworks such as Tensorflow has enabled optimizing computationally demanding models in many areas of physics and engineering (for example 
\cite{loper2014opendr, giftthaler2017automatic, de2018end, schenck2018spnets, kellman2019data, wang2020phase, bostan2020deep, zhou2020diffraction}) 
and also in optical imaging \cite{kellman2019data, vishniakou2020differentiable, ongie2020deep}.

This approach can be applied for adaptive optics \cite{vishniakou2020differentiable}.  Using a differentiable model of the optical system,  focusing in transmission through a single aberrating layer as well as in a reflection, epidetection configuration through two aberrating layers was achieved \cite{vishniakou2020differentiable}. 
Here, we extend this differentiable model-based adaptive optics method to scanning two-photon microscopy with a high numerical aperture objective as used for \textit{in vivo} imaging in brain tissue.

For this, a fluorescent guide star in an aberrated sample is probed with a small number of excitation modulations. 
The resulting dataset of pairs of excitation modulations and guide star images is sufficient for constraining model optimization in Tensorflow. 
This approach allows including detailed setup information in the optimization process and finds corrections independent of prior assumptions about the statistics of aberrations. We show experimentally that aberrations in a sample can be determined and corrected.

\section{Differentiable model-based approach for adaptive optics}

Modern machine learning frameworks efficiently implement model training by gradient based minimization of a loss function describing the mismatch between the output of a model and a target. They provide gradient-based optimizers and an automatic differentiation \cite{baydin2017automatic} framework, which leaves for the user only the task of implementing the desired model and loss function using a composition of differntiable functions. Here, we implement a model that simulates two-photon image formation in a scanning microscope depending on a set of free parameters which correspond to the sample aberration and are found through optimization (Fig. \ref{fig:setup}). An advantage of this approach is also the efficient implementation of optimization and model evaluation on a GPU, for example in Tensorflow, which was used here.

This approach is schematically illustrated in Fig. 1. First, a computational model is fitted to the two-photon microscope. This model accurately describes propagation of the laser beam through the microscope, including the spatial light modulator and the high numerical aperture objective, as well as image formation with the detected fluorescence. Then, aberrations are found by optimizing free model parameters corresponding to the undetermined sample aberrations (see Fig. 1 for details).

\section{Setup and image preprocessing}

The setup is schematically shown in Fig. \ref{fig:setup}. A custom built two-photon scanning microscope was used with components similar to the one described in \cite{vishniakou2020wavefront}. A laser beam (920 nm, Chameleon Discovery, Coherent) was expanded and reflected off a spatial light modulator (Meadowlark, HSP1920-1064-HSP8). The spatial light modulator was imaged onto the resonant scanning mirror with a pair of lenses. The scanning mirrors were imaged into the back focal plane of the objective and fluorescence was detected with a dichroic mirror and a lens focusing detected light onto a photomultiplier tube (Hamamatsu, H7422PA-40) using photon counting. The same excitation and fluorescence detection arrangement was used as described in \cite{vishniakou2020wavefront}, however, with a different objective (Nikon CFI Apo 60X W NIR,  N.A. 1.0, 2.8 mm W.D.) and correspondingly adjusted tube and scan lenses (all lenses used in the setup were achromatic doublets from Thorlabs). The microscope was controlled with Scanimage \cite{pologruto2003scanimage} which was integrated with custom software written in Python for SLM control and for synchronizing probing modulations with image acquisition using Scanimage API callbacks.

Fluorescence images were recorded using photon counting at $512\times512$ pixels resolution at 30 Hz. This resulted in sparse images with low counts of photons per pixel, with many discontinuities (gaps) in intensities hindering the correct estimation of the similarity with model output. Therefore all fluorescence images were preprocessed with a low-pass filter: a discrete Fourier transform was applied to the images and frequencies exceeding $0.15$ of the pattern resolution were discarded before inverse transformation. Examples of images before and after preprocessing are shown in Fig. \ref{fig:model_fitting} (columns two and three, respectively).

\section{Computational modeling and optimization}

\subsection{Computational model of the optical setup}

\begin{figure}
    \centering
    \includegraphics[width=0.85\textwidth,trim={5cm 3cm 10cm 0},clip]{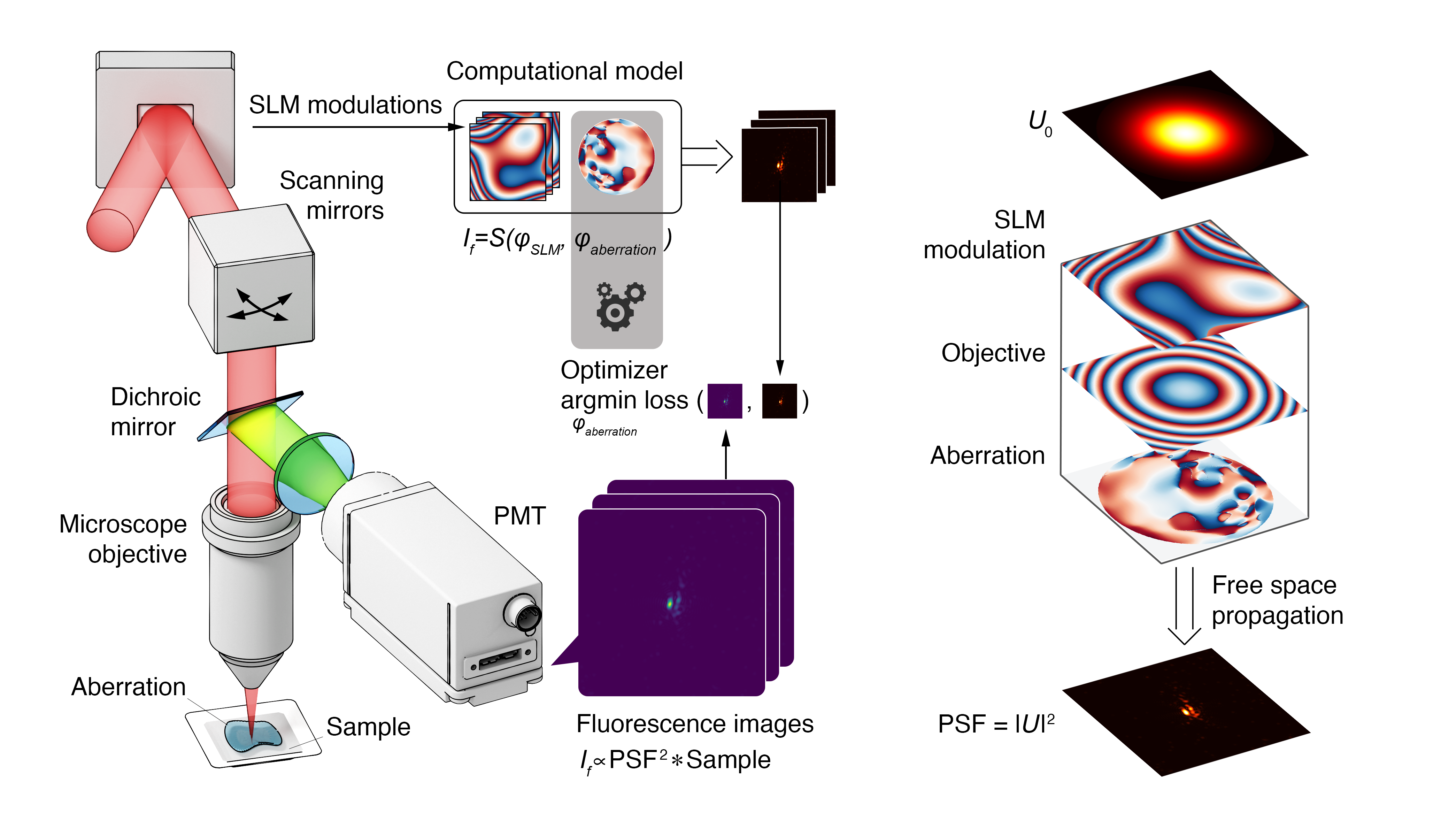}
    \caption{Left: Schematic of the two-photon microscope and corresponding model optimization. A laser beam is scanned with scanning mirrors and focused through an objective. Fluorescence light is collected with a dichroic mirror onto a photomultiplier tube. A computational model is set up to describe the microscope in the absence of aberrations. After introducing aberrations, the computational model receives phase modulations displayed on the SLM as input. The model output is a set of point spread function (PSF) images that are compared with experimentally recorded PSF images through the loss function. The optimizer (running in Tensorflow on a GPU) finds a sample aberration (and corresponding correction) by matching of the model output with experimental PSF images.
    Right: The computational model is formulated in expression (\ref{eq:simulation}). The beam cross-section, represented as a complex amplitude, is multiplied by phase objects representing SLM probing modulation, objective, and aberration, respectively, and is propagated to the focal plane of the objective using the angular spectrum method to calculate the resulting PSF. The two-photon image is the convolution of the squared PSF intensity and the object function, resulting in just the point spread function squared for a guide star. The beam profile $U_0$ and objective parameters are adjusted to describe the experimental setup. The SLM modulation is the model input, the PSF is the model output, the aberration is found through optimization.}
    \label{fig:setup}
\end{figure}

To implement differentiable model-based adaptive optics for two-photon microscopy, first a differentiable model of the setup needs to be established. The main elements of the setup model are illustrated in Fig.\ref{fig:setup}, right side. The model consists of, first, a phase modulation describing the spatial light modulator (SLM), second, a phase function describing the focusing objective,  and, third, a (unknown) sample phase aberration. The optimization process is schematically illustrated in Fig.\ref{fig:setup} (see figure legend for details).

The computational model is based on Fourier optics. Light propagation along the optical axis ($z$-axis; x, y, z are spatial coordinates) is represented as a complex amplitude $U(x, y, z)$. The wavefront propagates in free space and through a sequence of planar phase objects along the optical axis. The interaction of the wavefront $U(x, y, d)$  with a phase object $\phi(x, y, d)$ at plane $d$ is described as a multiplication:
\begin{equation}\label{eq:phase_interaction}
U(x, y, d)\cdot\exp\left[{i\phi(x,y,d)}\right].
\end{equation}
Free space propagation of the wavefront over a distance $d$ is calculated using the angular spectrum method with the following operator \cite{goodman2005introduction}:

\begin{equation}\begin{aligned}\label{eq:propagation_operator}
U(x, y, z+d) &= P_d(U(x, y, z)) = \iint A(f_x, f_y; z)\,\mathrm{circ}\left(\sqrt{(\lambda f_X)^2+(\lambda f_Y)^2}\right)\\&\times H\exp\left[i2\pi(f_Xx+f_Yy)\right] \,\mathrm{d}f_X\,\mathrm{d}f_Y.
\end{aligned}\end{equation}
Here, $A(f_X, f_Y; z)$ is the Fourier transform of $U(x, y, z)$, $f_X$ and $f_Y$ are spatial frequencies, the circ function is 1 inside the circle with the radius in the argument and 0 outside \cite{goodman2005introduction},
and $H(f_X, f_Y) = \exp\left[i2\pi\frac{d}{\lambda}\sqrt{1-(\lambda f_X)^2-(\lambda f_Y)^2}\right]$ is the optical transfer function. 
Light intensity as measured at the sensor 
is given by 
\begin{equation}\label{eq:intensity}
I(x, y, z) = \left|U(x, y, z)\right|^2.
\end{equation}
For two-photon imaging \cite{denk1990two} the induced fluorescence intensity is proportional to the square of the excitation intensity: 
\begin{equation}\label{eq:fluorescence_intensity}
I_{\mathrm{f}} \propto I^2 = \left|U\right|^4.
\end{equation}

Using these equations, the image of a fluorescent bead in the microscope (or the PSF) is simulated with the following function,  taking into account that the propagating wavefront is modulated by the SLM and  phase aberration of the sample:
\begin{equation}\begin{aligned}\label{eq:simulation}
I_{\mathrm{f}} = S(\phi_\mathrm{SLM}, \phi_\mathrm{aberration}) &= \left| P_{f_\mathrm{MO}}(U_{0}\cdot\exp\left[i(\phi_\mathrm{SLM}+\phi_\mathrm{MO}+\phi_\mathrm{aberration}) \right])\right|^4,
\end{aligned}\end{equation}
where $\phi_\mathrm{SLM}$ is the SLM phase modulation, $\phi_\mathrm{aberration}$ is the phase surface of the aberration, $U_{0}$ is the beam cross-section (complex) amplitude, and $f_\mathrm{MO}$ and $\phi_\mathrm{MO}$ are the focal length and the corresponding phase modulation of the microscope objective, respectively. 
The SLM modulation is modelled directly at the same z-plane as the objective,
which allows omitting nonessential computations, since it is imaged onto the back focal plane of the microscope objective.
The aberration is also modelled at the same plane, saving computations and simplifying the correction: since the phase function of the aberration is found at the SLM plane, its inverse can be directly applied to the SLM without any additional computations of propagation between different planes.

\subsection{Fitting the computational model to the experimental setup}

The goal of the optimization process is matching the computational model to experimental observations with correlations as a similarity measure.  The model images therefore have to match in each pixel with the images observed in the microscope in the absence of sample aberrations. When responses to probing SLM modulations are measured under these aberration-free conditions, model optimization is expected to yield a flat aberration. Optimization without sample aberrations therefore can be used for validating the correspondence between the model and the microscope.
The model equation (\ref{eq:simulation}) contains fixed parameters $U_{0}$ (Gaussian amplitude profile with flat phase, $\sigma=2.85$ mm), $\lambda=920$ nm, $f_\mathrm{MO}=1700$ mm, and corresponding $\phi_\mathrm{MO}$ which were matched to the experimental setup by taking into account pupil sizes $d=4.0$ mm, field of view size ($60 \times 40 \mu$m at zoom level 10), and the resolution of imaging and simulation ($512\times512$ pixels). Significant elongation of $f_\mathrm{MO}$ allows approximate lateral upscaling of simulated images for matching the microscope zoom level without resampling of the angular spectrum. Additionally, minor rotational misalignments of the SLM in the image plane were observed and adjusted for by corresponding counter-rotations of the recorded fluorescence images, since such rotations could not be corrected by adjusting the SLM phase. 
A fixed zoom level and field of view were used for all experiments and the model was adjusted for these conditions. 
The parameters were tuned in the following procedure: first, a sparse aberration-free fluorescent guide star (1 $\mu$m diameter fluorescent beads embedded in 3 $\%$ agarose) was imaged under various modulations (Zernike modes Z1 through Z10 displayed one-by-one with alternating magnitudes). The parameters were then manually tuned to closely match computed to actual images. For this, zoom level and SLM orientation were varied while displaying a set of different SLM phase modulations.  This procedure needed only to be done once and was therefore performed manually, but a computational optimization would also be possible. The resulting model was validated by running the model optimization for finding corrections in the absence of introduced sample aberrations. After successful parameter optimization, the residual aberrations of the system were close to zero, indicating that the computational model accurately described the experimental microscope setup. 
As shown in Fig. \ref{fig:model_fitting}, after tuning of the model parameters, good correspondence between experimentally observed and computationally generated patterns was achieved (shown here after optimization in an aberrating sample, see below).

\subsection{Model optimization and loss function}
\begin{figure}
    \centering
    \includegraphics[width=0.7\textwidth,trim={0 0 0 0},clip]{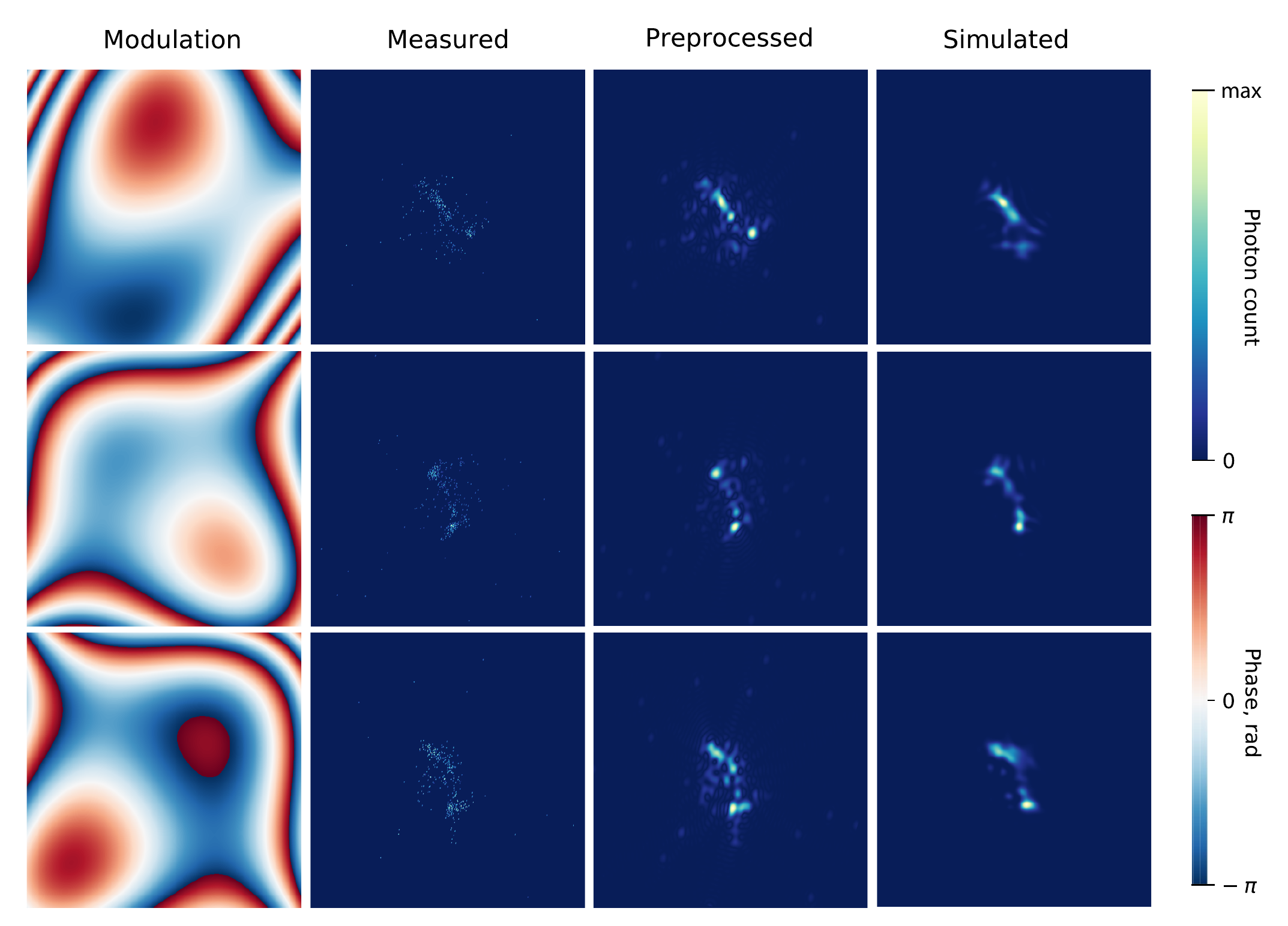}
    \caption{Correspondence between computationally generated and experimentally measured images of fluorescent beads. Three representative examples (rows), recorded in an aberrating sample: probing modulations displayed on SLM (first column), and corresponding PSF images as measured with photon counting (second column), preprocessed for model fitting (third column) and simulated with the optimized model. Field of view size is $40\times40 \mu$m.}
    \label{fig:model_fitting}
\end{figure}

Sample aberrations introduce an unknown phase function into the optical setup. To mirror this situation computationally, a phase surface is added as a set of free parameters to the model \cite{vishniakou2020differentiable}. This unknown phase surface needs to be adjusted through optimization in such a way that it describes the introduced sample aberration. 
After successful optimization,  the aberration phase surface is known, as verified by the model again matching the optical setup (Fig. \ref{fig:model_fitting}), and can therefore be corrected.

We found that optimization with a single image of the fluorescent bead and corresponding single SLM phase modulation did often not result in satisfactory results since multiple possible phase modulations can generate similar planar PSF images. To constrain the optimization process, we therefore probed the guide star with a set of different excitation modulations as in \cite{vishniakou2020differentiable} by displaying randomly generated phase patterns on the SLM and imaging the resulting changes in the aberrated guide star. 

In total, 20 such pairs of SLM phase aberrations and corresponding two-photon images served as the input to the computational model. Images were recorded using photon counting at a frame rate of 30 Hz. (The SLM has a maximum frame rate of more than 500 Hz, but the control loop implemented here ran only at 2 Hz since SLM and image acquisition were not closely integrated.) Probing phase modulations were generated by summing Zernike modes Z1 through Z10 with coefficients drawn form a uniform random distribution in the range of $-1$ to $1$ and were displayed on the SLM while corresponding fluorescence images $I_{\mathrm{f}}$ were recorded.

The unknown aberration phase function was then found in an optimization process with the goal of matching measured and simulated images:
\begin{equation}\label{eq:argmin}
\argmin_{\phi_{\mathrm{aberration}}} \sum_{j=1}^{N}\mathrm{loss}(S(\phi_{\mathrm{SLM}_{j}}, \phi_{\mathrm{aberration}}), I_{\mathrm{f}_{j}}).
\end{equation}
Here, $\mathrm{SLM}_j$ and $\phi_{\mathrm{aberration}_j}$ are pairs of SLM probing modulations and corresponding fluorescence images recorded in the two-photon microscope. $S$ is the microscope model as specified in (\ref{eq:simulation}). The loss function is defined to reflect the similarity between simulated and measured fluorescent images:
\begin{equation}\label{eq:loss}
    \mathrm{loss}(\mathrm{prediction}, \mathrm{target}) = -r\left[\mathrm{prediction}, \mathrm{target}\right]\cdot\ln{(\sum^\mathrm{pixels}|\mathrm{prediction}|)},
\end{equation}
where $r\left[\mathrm{prediction}, \mathrm{target}\right]$ is Pearson's correlation coefficient, and $\ln{(\sum^{\mathrm{pixels}}|\mathrm{prediction}|)}$ is an additional cost factor for light intensity conservation. 
Since Pearson's correlation coefficient is not sensitive to the magnitude of the prediction, the optimizer likely converges to a solution which provides high correlation, but discards some of the light by redirecting it out of the field of view, and therefore usually does not result in good corrections. Therefore, additionally introducing the sum of the total intensity promotes solutions that do not discard excitation intensity. The logarithm ensures that the slope of this regularization factor is always smaller than that of the correlation coefficient, making it a secondary optimization goal.

The model (\ref{eq:simulation}) was implemented using Tensorflow 2.4 \cite{tensorflow2015-whitepaper}, adapting the angular spectrum method from \cite{diffractio} and using the optimization algorithm Adam with a learning rate of $0.01$. According to expression (\ref{eq:simulation}), the phase $\phi_{\mathrm{aberration}}$ is represented as a real-valued tensor, which is a requirement for optimization variables in Tensorflow. All modulations and sample responses were packed in a single batch and used all at once in each of the optimizer's iterations. 1000 optimization steps were typically sufficient for  reaching a correlation coefficient between model and observations of $>0.9$. The optimization took between 1 and 2 minutes on a workstation with four Nvidia Titan RTX GPUs used in data parallel mode. The optimization was terminated when the correlation coefficient between computational model and experimental data plateaued. The resulting aberration phase function was negated (multiplied by $-1$) for displaying on the SLM and in this way results in the correction of the aberration. 

\section{Two-photon imaging through aberrations}
 
To test the approach experimentally, a sample aberration was introduced by covering the fluorescent beads with a layer of vacuum grease on a microscope cover slide. For correcting the induced aberrations, images were taken with a single guide star at the center of a field of view with dimensions of $60\times40 \mu$m. 

\begin{figure}
    \centering
    \includegraphics[width=0.7\textwidth,trim={0 0 0 0},clip]{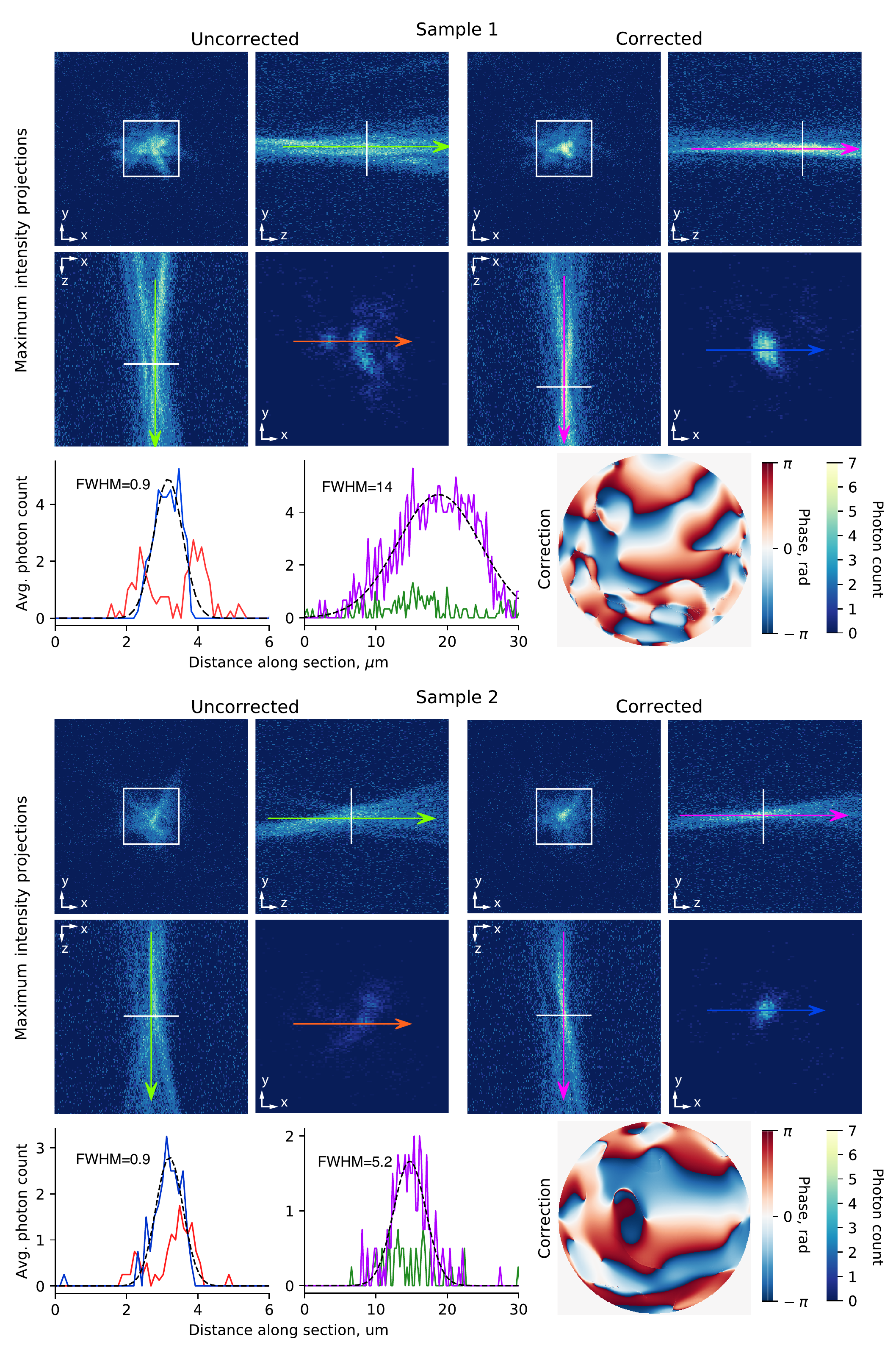}
    \caption{Two representative examples of two-photon volume images of a $1 \mu$m diameter fluorescent bead embedded in agarose before and after correction. Sample 1, top row, left: Maximum intensity projection of volume image of a fluorescent bead in axial (z) direction (see bottom row for color bar). Projections along x- and y-axes are shown to the right (first row, left center) and below (second row, left). Second row, left center: slice at maximum intensity, indicated by white lines in x- and y-projections averaged over 4 frames recorded with $0.25 \mu$m axial spacing. Image is zoomed-in as indicated by white box in top left image, size is $10\times10 \mu$m. Top row, center right: same as left side after correction, again with corresponding x- and y- projections. Second row, right: slice at maximum intensity, indicated by white lines in x- and y-projections averaged over 4 frames recorded with $0.25 \mu$m axial spacing after correction. Third row, left: lateral cross-sections before and after correction as indicated by correspondingly colored arrows in row two, including Gaussian fit of corrected profile (black dashed line). Third row, center: axial cross sections along the correspondingly colored arrows in the second row including Gaussian fit of corrected image. Third row, right: aberration correction. Sample 2: as described for sample 1.}
    \label{fig:corrections}
\end{figure}

Two representative results are shown in Fig. \ref{fig:corrections}. Three orthogonal maximum intensity projections, before and after correction, of a fluorescent bead recorded in a volume with axial step size of 0.25 $\mu$m between planes are displayed. The third and sixth rows show the improved intensity profiles along representative cross-sections through the fluorescent bead in lateral and axial directions, respectively.  Corrections resulted in an increase in intensity and an improved focus with a full width at half maximum (FWHM) after correction of $0.9 \mu$m for both samples, and an axial FWHM of $14 \mu$m and $5.2 \mu$m, respectively (measured with a $1 \mu$m diameter bead). The intensity profiles before correction could not be fitted with  Gaussians. The slices at the focal plane (at maximum intensity)  were averaged over four frames (axially spaced by 0.25$\mu$m). Axial profiles were averaged over $3\times3$ pixels around the center of the focus (maximum intensity) and averaged over the lateral dimensions.

\section{Discussion and conclusions}
 
We developed a differentiable model-based approach for adaptive optics for two-photon scanning microscopy. The method takes advantage of combining a differentiable model of the known microscope and unknown sample with the optimization techniques implemented in machine learning frameworks \cite{vishniakou2020differentiable}. We show that, with an appropriate cost function (equation \ref{eq:loss}), a small number of probing modulations (recorded at high frame rates and with low photon count rates as often observed when imaging \textit{in vivo}) is sufficient for finding sample aberrations through model optimization and to correct them.

Different from optimization approaches that directly aim to optimize focus intensity, the first optimization objective here was matching the focus shape of the model to the one observed in the microscope. However, we added an additional, secondary term to the cost function to also optimize image intensity directly.

A limitation of the current implementation for dynamic samples is the required correction time. Here, we used 20 probing modulations recorded at a frame rate of 30 Hz, thus requiring a minimum of  0.66 seconds for data acquisition. Model optimization took between 1 and 2 minutes, which is ultimately limited by the computational speed of the GPU and could be accelerated with improved or more GPUs.

As seen in Fig. 3, the focus after optimization did in particular in axial direction not reach the diffraction limit (as measured with 1 $\mu$m diameter fluorescent beads). While such axially extended point spread functions are often used for \textit{in vivo} imaging (see for example \cite{song2021neural} for a recent discussion), the quality of the corrections could be improved in several ways. 
A limitation of the current data, in particular for correcting higher order aberrations, is their limited dynamic range. 
The dynamic range could  for example be extended by combining multiple fluorescence images recorded with different integration times \cite{vinegoni2018high}.  
Additional improvements in performance are expected from using a model approximation that is more accurate for high numerical aperture objectives  than the one used here\cite{Thao2020phase}. Together, such improvements will more accurately reflect the physical setup and aberrations and therefore improve optimization results. 
As an alternative to using multiple modulations recorded in the same focal plane, also axial slices or an entire volume could be used for optimization.

Compared with other optimization approaches that rely on iterative methods, including prior information about the experimental setup additionally constrains the optimization process \cite{yang2015model}. Compared to neural network approaches \cite{angel1990adaptive, paine2018machine, swanson2018wavefront, andersen2019neural, andersen2020image, jin2018machine, hu2019learning, cheng2019artificial, vishniakou2020wavefront, saha2020practical}, differentiable model-based approaches have the advantage that they don’t rely on a predetermined model of sample aberrations. 

Overall, the presented approach, similar to other iterative approaches, can be used for correcting aberrations that are approximately stationary, and is under these conditions compatible with imaging in biological samples. 

\section*{Funding}
Max Planck Society, caesar

 
\section*{Disclosures}
The authors declare no conflicts of interest.

\bibliographystyle{unsrt}  
\bibliography{references}  


\end{document}